\documentstyle[aps,prb,eqsecnum,epsf]{revtex}
\begin{document}
\title{Edge reconstructions in fractional quantum Hall systems}
\author{Yogesh N. Joglekar$^{1,2}$, Hoang K. Nguyen$^{1}$, and Ganpathy Murthy$^{1}$}
\address{$^{1}$Department of Physics and Astronomy, University of Kentucky, Lexington, KY 40506, \\
$^{2}$ Theoretical Division, Los Alamos National Laboratory, Los Alamos, NM 87544.}
\maketitle
\begin{abstract}
Two dimensional electron systems exhibiting the fractional quantum Hall effects are characterized by a quantized 
Hall conductance and a dissipationless bulk. The transport in these systems occurs only at the edges of the 
incompressible quantum Hall regions, where gapless excitations are present. We present a {\it microscopic} 
calculation of the edge states in the fractional quantum Hall systems at various filling factors using the 
extended Hamiltonian theory of the 
fractional quantum Hall effect. We find that at $\nu=1/3$ the quantum Hall edge undergoes a reconstruction 
as the background potential softens, whereas quantum Hall edges at higher filling factors, such as 
$\nu=2/5, 3/7$, are robust against reconstruction. We present the results for the 
dependence of the edge states on various system parameters such as temperature, functional form and range of 
electron-electron interactions, and the confining potential. Our results have implications for the tunneling 
experiments into the edge of a fractional quantum Hall system.
\end{abstract}


\section{Introduction}
\label{sec: intro}
The physics of two-dimensional (2D) electron systems in a strong magnetic field has lead to the exploration 
and understanding of various strongly correlated states in the fractional quantum Hall 
regime.~\cite{fqhereview} The fractional quantum Hall effects (FQHE) at filling factors $\nu=1/(2p+1)$ were 
explained by Laughlin, using trial wavefunctions which lead to incompressible states with quasiparticle 
excitations having fractional charge $e^{*}=e/(2p+1)$. Jain obtained a unified description of all principal 
filling factors $\nu=p/(2ps+1)$ by using the composite fermion picture, where each electron is dressed 
with $2s$ units of statistical flux and forms a composite fermion (CF).~\cite{jain} At mean-field level, the 
composite fermions experience a uniform reduced magnetic field, $B^{*}=B/(2sp+1)$, and fill an integer 
number $p$ of CF Landau levels. This approach, based on trial wavefunctions and numerical 
diagonalization, has been very successful in describing various properties of fractional quantum Hall 
states; however, it does not allow calculation of dynamic response functions and suffers from the inevitable 
finite-system-size limitations. These limitations are overcome by an approach based on the Chern-Simons 
transformation, which attaches flux to the electrons via a Chern-Simons gauge field, effectively transforming 
the fermions 
into (hard-core) bosons or fermions depending on the number of flux quanta attached.~\cite{csreviews} This 
approach does start with a microscopic Hamiltonian, but after the Chern-Simon transformation it {\it does 
not} retain the lowest Landau level limit, and does not distinguish clearly between the high-energy 
(cyclotron) and the low-energy (guiding center) degrees of freedom. Most recently, the extended Hamiltonian 
theory developed by Shankar and Murthy~\cite{hamreview} has been successful in obtaining a unified 
picture of the FQHE at various fillings. This approach has the advantage that it starts with a microscopic 
Hamiltonian, permits analytical calculations in the thermodynamic limit, and it also retains the 
lowest-Landau-level limit. This approach has also been applied successfully to study states with broken 
translational symmetries, such as a Wigner crystal of composite fermions.~\cite{narevich}
	
In an {\it ideal} quantum Hall system, the bulk is uniformly incompressible and gapless excitations are 
available only at the edges leading to transport only at the edges as well. In experimental samples 
ubiquitous disorder leads to islands of incompressible quantum Hall fluid embedded in a compressible 
electronic background; the global Hall quantization occurs when these incompressible islands percolate. 
In either case, the transport occurs only the edges of incompressible regions.~\cite{yacobi,zhitenev} 
The dynamics of the fractional quantum Hall edge states was first studied by 
Wen~\cite{wen} who proposed that the edges behave effectively as Chiral Luttinger liquids (CLL). Later it was 
discovered that, due to electrostatic and exchange considerations, the edge of a fractional quantum Hall 
fluid divides into strips of compressible and incompressible fluids.~\cite{chlovskii,allan} In particular, 
the edge of a quantum Hall fluid at filling factor $\nu=1$ undergoes a reconstruction, where a strip of charge 
about a magnetic-length wide separates away from the bulk as the confining potential created by the 
neutralizing background charge density softens.~\cite{allan,chamon} The reconstructed edge consists of three 
counter-propagating channels; its dynamics are qualitatively different from those of a single CLL edge. This 
led to work on fractional edge-states using the composite fermion picture~\cite{brey,chl,eric} and 
the effective-field-theory models which included the effects of disorder, multiple edge channels, and 
long-ranged interactions.~\cite{kane,ana,ulrich,levitov,goldman} The first microscopic calculation of the 
edge-states at filling factor $\nu=1/3$, using exact diagonalization, showed that the edge of the Laughlin 
liquid also undergoes a reconstruction as the confining potential is softened.~\cite{wan} However, 
there are various issues which neither the effective-field-theory models nor exact diagonalization methods 
address conclusively. For example, the number of edge modes present at various filling factors and their 
dependence on the system parameters is not clear. The question of whether long-ranged interactions are 
{\it necessary} for edge reconstruction is not clearly resolved, although the necessity has been implicitly 
assumed in the literature.~\cite{chamon,wan,wan2} The phenomenological models of the edge-states provide 
various concrete predictions for the $I(V)$ characteristics of tunneling into the edge of the 
sample.~\cite{wen,kane,levitov} These predictions depend on the number and the nature of the edge channels. 
Recent experiments on tunneling into the edge of a fractional quantum Hall fluid show discrepancies between 
the theoretical predictions and the observed characteristics.~\cite{goldman,mandal,dhlee,kun,grayson,chang} It 
is, therefore, crucial to have a microscopic understanding of the number of edge modes in the fractional 
quantum Hall systems at various filling factors and their dependence on the system parameters.  

In this paper, we use the extended Hamiltonian theory to the study nature of edge states, and address some 
of the unresolved issues. This method is applicable at any principal filling factor and provides a tool 
for the microscopic analysis of the edge states. The outline of the paper is as follows: In the next 
section, we recall the essentials of the Hamiltonian theory and present the Hartree Fock mean-field 
approximation for the edge states. Section~\ref{sec: results} contains results for the edge 
reconstructions at filling factors $\nu=1/3$, and $\nu=2/5,3/7$ for various system parameters, such as 
temperature, nature of interactions, and the range of the confining potential. Edge reconstruction is a 
purely quantum phenomenon. In a classical system, the electron density profile near the edge would follow 
the background charge profile to minimize the Hartree (electrostatic) energy cost. However, in a quantum 
treatment, the electron density deviates from the background charge density to take advantage of the 
exchange interactions, which lower the total energy of the system. The ratio of the effective exchange term 
and the electrostatic (Hartree) term, and consequently the possibility of edge reconstruction, naturally 
depends on the microscopic interaction. In this section we also explore this dependence by using different 
functional forms for the microscopic interaction. Section~\ref{sec: remarks} contains concluding remarks. 


\section{Hamiltonian Theory and Mean-field Approximations}
\label{sec: ham}
We start this section with the basics of the extended Hamiltonian theory.~\cite{hamreview} The fundamental 
difficulty in developing analytical approximations starting from a microscopic Hamiltonian for the FQHE is 
that there are more single-particle states available in the lowest Landau level than there are electrons. In 
the Extended Hamiltonian theory, this macroscopic degeneracy is removed by introducing extra degrees of 
freedom called the pseudo-vortices, which leads to an enlargement of the Hilbert space of the problem. This 
enlarged space is the space of CFs (composites of electrons and pseudo-vortices) in a reduced 
magnetic field $B^{*}=B/(2p+1)$, where the composite fermions fill the first $p$ CF Landau levels. Thus the 
Hartree-Fock state of composite fermions with the first $p$ Landau levels filled provides a 
{\it non-degenerate} starting point for further systematic approximations, thereby eliminating the degeneracy 
problem. However, since we have introduced extra degrees of freedom, namely the pseudo-vortices, we then 
impose an equal number of constraints so that the dimensions of the physical Hilbert space are maintained. 
To this end we introduce two sets of guiding-center coordinates 
\begin{eqnarray}
\label{eq: ham1}
\vec{R_e}& = &\vec{r}-\frac{l^2}{1+c}\hat{z}\times\vec{\Pi}, \\
\label{eq: ham2}
\vec{R_v}& = &\vec{r}+\frac{l^2}{c(1+c)}\hat{z}\times\vec{\Pi},
\end{eqnarray}
where $c^2=2\nu$, and $\vec{r}$ and $\vec{\Pi}$ are the position and velocity operators of the composite 
fermions. The electron guiding center $\vec{R_e}$ and the pseudo-vortex guiding center $\vec{R_v}$ commute 
with each other and satisfy the algebra
\begin{eqnarray}
\label{eq: ham3}
[R_{e\alpha},R_{e\beta}]=-il^2\epsilon_{\alpha\beta},\\
\label{eq: ham4}
[R_{v\alpha},R_{v\beta}]=+i\frac{l^2}{c^2}\epsilon_{\alpha\beta}.
\end{eqnarray}
Thus, the electron guiding-center coordinates satisfy the magnetic algebra with charge $-e$, whereas the 
pseudo-vortex guiding-center algebra represents an object with charge $+ec^2$. To calculate the matrix 
elements of these operators in the composite-fermions basis, we use the single-particle states of the 
composite fermions in the reduced effective field $B^{*}$. In the Landau gauge, a single-particle state 
$|nX\rangle$ is characterized by the Landau level index $n$ and the guiding-center coordinate $X$. In the 
real-space representation, the (un-normalized) single-particle wavefunction is
\begin{equation}
\label{eq: ham}
\langle\vec{r}|nX\rangle=e^{iXy/l^{*^2}}e^{-(x-X)^2/2l^{*^2}}H_n\left[(x-X)/l^{*}\right],
\end{equation}
where $H_n(x)$ are the Hermite polynomials and $l^{*}=\sqrt{h/eB^{*}}$ is the magnetic length in the reduced 
field $B^{*}$ seen by the composite fermions. Using this basis, it is straightforward to express the 
electron density $\rho_e$ and the pseudo-vortex density $\rho_v$ operators in second-quantized notation:
\begin{eqnarray}
\label{eq: ham5}
\rho_e(\vec{q})& = & e^{-q^2l^{*^2}(1-c^2)/4}\sum_{n_i,X_i}d^{\dagger}_{n_1X_1}d_{n_2X_2}\langle n_1X_1|
	e^{-i\vec{q}\cdot\vec{R_e}}|n_2X_2\rangle, \\
\label{eq: ham6}
\rho_v(\vec{q})& = & e^{-q^2l^{*^2}(1-c^2)/4}\sum_{n_i,X_i}d^{\dagger}_{n_1X_1}d_{n_2X_2}\langle n_1X_1|
	e^{-i\vec{q}\cdot\vec{R_v}}|n_2X_2\rangle.
\end{eqnarray}
Here $d^{\dagger}_{nX}$ ($d_{nX}$) are the CF creation (annihilation) operators and the exponential 
prefactor $e^{-q^2l^{*^2}(1-c^2)/4}$ arises from projecting the cyclotron coordinate on to the 
lowest Landau level.~\cite{hamreview} We have now expressed the electron density operators, and hence the 
microscopic Hamiltonian for the fractional quantum Hall systems, in terms of composite fermion operators. 
Therefore the original problem of interacting electrons in the {\it fractional} quantum Hall regime is 
converted into a problem of 
composite fermions in the {\it integer} quantum Hall regime where various many-body approximations are 
applicable. This mapping, of course, must by accompanied by the constraint that the pseudo-vortex degrees of 
freedom are not physical; in other words, physics remains unchanged if 
$\rho_e\rightarrow \rho_e+\alpha\rho_v$ for arbitrary $\alpha$. This constraint is, by definition, trivially 
satisfied in the exact solution of the problem. For a 
uniform case, it was shown by Murthy~\cite{conserv} that the (time-dependent) Hartree-Fock approximation 
respects this constraint. Therefore Hartree-Fock mean-field approximation is a particularly good starting 
point~\cite{conserv,read} for the calculation of response functions.

\begin{figure}[t]
\begin{center}
\hspace{2cm}
\epsfxsize=5in
\epsffile{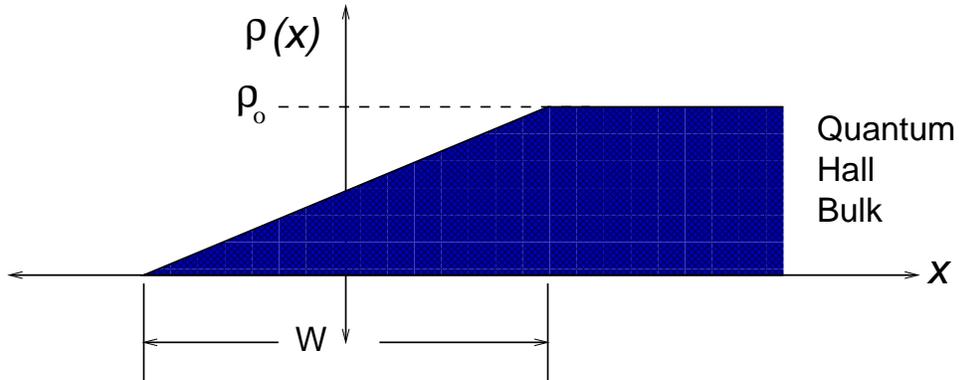}
\caption{Neutralizing background charge-density profile: The background charge density vanishes linearly 
over a width $W$ around the edge at $x=0$ and is given by 
$\rho_{bg}(x)=-\frac{\rho_0}{2}\left[1+\frac{2x}{W}\right]$ for $|x|<W/2$. Here $\rho_0$ is the bulk 
composite-fermion density. We find that increasing $W$ leads to edge-reconstructions.}
\label{fig: bg}
\end{center}
\end{figure}

Now let us apply this formalism to the edge problem. We assume that the sample-edge runs along $x=0$ 
breaking the translation invariance along the $x$-axis. Since the system is translationally invariant along 
$y$-axis, it is advantageous to use the Landau gauge single-particle states with guiding centers along the 
$x$-axis. The Hamiltonian of our system consists of two terms: the electron-electron repulsion $H_e$ and the 
electron-background attraction $H_b$,
\begin{eqnarray}
\label{eq: ham7}
H_e & = &\frac{1}{2A}\sum_{\vec{q}}V_{ee}(\vec{q}):\rho_e(\vec{q})\rho_e(-\vec{q}):\\
\label{eq: ham8}
H_b & = &\sum_{\vec{q}}{V}_{eb}(-\vec{q})\rho_e(\vec{q})
\end{eqnarray}
where $V_{ee}(\vec{q})$ is the electron-electron interaction, and $V_{eb}$ is the one-body potential created 
by the background charge-density $\rho_b(x)$ which is non-uniform near the edge $x=0$,
\begin{equation}
\label{eq: ham9}
V_{eb}(x) =  \int^{\infty}_{-\infty} dx'\, V_{eb}(x-x')\rho_b(x'). 
\end{equation}
Notice that since the interaction Hamiltonian (\ref{eq: ham7}) and the background Hamiltonian 
(\ref{eq: ham8}) are expressed {\it only} in terms of the electron-density operators, it is possible to 
recast the Hamiltonian, using Eq.(\ref{eq: ham5}), entirely in terms of CF creation and annihilation 
operators. We consider a background charge-density which vanishes linearly over width $W$: $\rho_b(x)=0$ for 
$x<0$, $\rho_b(x)=-\frac{\rho_0}{2}\left[1+\frac{2x}{W}\right]$ for $|x|<W/2$, and $\rho_b(x)=\rho_0$ for 
$x>W/2$ where $\rho_0$ is the CF density in the bulk (See Fig.~\ref{fig: bg}). In general, the functional forms of 
the electron-electron and electron-background interactions may be different; for example, if 
the neutralizing background layer is a distance $d$ away from the 2D electron layer, for Coulomb repulsion, 
we will have $V_{ee}(q)=2\pi e^2/q$ and $V_{eb}(q)=-V_{ee}(q) e^{-qd}$. 

The standard Hartree-Fock decoupling~\cite{negele} of the interaction term (\ref{eq: ham7}) leads to the 
following mean-field Hamiltonian,
\begin{eqnarray}
H_{HF} & =& \sum_{n_iX_i}\langle n_1X_1|\hat{U}_{HF}+\hat{V}_{eb}|n_2X_2\rangle 
d^{\dagger}_{n_1X_1}d_{n_2X_2},\\
\label{eq: ham10}
\langle n_1X_1|\hat{U}_{HF}|n_2X_2\rangle &=& \sum_{n_iX_i}\langle n_1X_1,n_3X_3|\hat{V}^A_{ee}|n_2X_2,n_4X_4
\rangle\sigma_{n_3n_4}(X_3,X_4),
\end{eqnarray}
where $V^A_{ee}$ are the anti-symmetrized matrix elements for the electron-electron interaction and 
$\sigma_{n_3n_4}(X_3,X_4)=\langle d^{\dagger}_{n_3X_3}d_{n_4X_4}\rangle$ is the density matrix determined 
self-consistently. We introduce a finite temperature (much smaller than the bulk-gap between the CF Landau 
levels for a given interaction~\cite{conserv}) to achieve rapid convergence of the iterations which 
determine the density matrix $\sigma$. Although this leads to non-integer CF occupation numbers, we have 
checked that the results presented evolve continuously to integer CF occupation numbers as the temperature 
is lowered. We consider the mean-field solutions which are uniform along the 
$y$-axis, $\sigma_{n_3n_4}(X_3,X_4)=\sigma_{n_3n_4}(X_3)\delta_{X_3X_4}$. Since the Landau-gauge basis 
quantum numbers are good quantum numbers in the bulk, it is clear that the mixing between different CF 
Landau levels occurs only because of the non-uniform background charge density and is confined to a region 
near the edge. In the following section, we present the results of the Hartree-Fock mean-field theory for 
filling fractions $\nu=1/3$ and $\nu=2/5,3/7$. We assume that the electron-electron and the 
electron-background interactions differ only by a sign, $V_{ee}(q)=-V_{eb}(q)$ (except when we compare the 
results with those 
of Wan {\it et al.}~\cite{wan}). To investigate the effect of the nature of interactions on the edge 
dynamics, we consider Gaussian interaction, $V_{ee}(q)=V_0\exp(-\alpha q^2l^{*^2}/2)$, characterized by 
dimensionless real-space range $\alpha$ and screened Coulomb interaction, 
$V_{ee}(q)=2\pi e^2/\sqrt{q^2+q_{TF}^2}$, which becomes truly long-ranged as $q_{TF}l^{*}\rightarrow 0$.


\section{Results}
\label{sec: results}
Let us start this section with results for filling factor $\nu=1/3$. In the Hamiltonian formalism, the 
$\nu=1/3$ system is mapped onto a CF system at filling factor $p=1$. In fact, if we take into account only 
the occupied CF Landau level and neglect the higher CF Landau levels since they are gapped in the bulk, this 
system is identical to the $p=1$ quantum Hall system studied by Chamon and Wen.~\cite{wen} It is therefore 
not surprising that the quantum Hall edge at $\nu=1/3$ undergoes an edge reconstruction as the width $W$ of 
the background charge-density profile increases.~\cite{wan} However, it is crucial to check the stability of 
this reconstruction when higher CF Landau levels are included (in an exact treatment, one must take into 
account {\it all} CF Landau levels).

\begin{figure}[t]
\begin{center}
\begin{minipage}{20cm}
\begin{minipage}{9cm}
\epsfxsize=3.3in
\epsffile{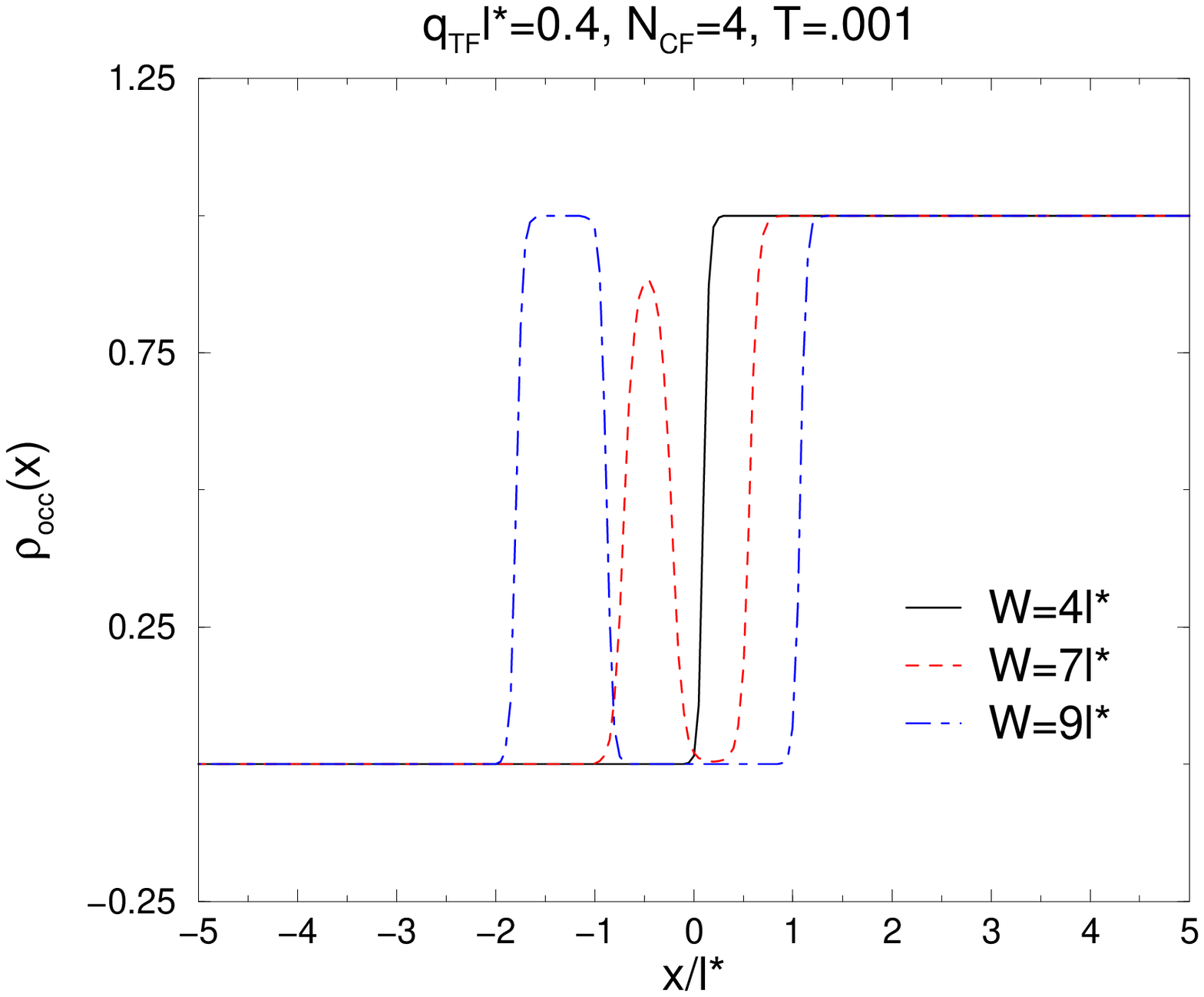}
\end{minipage}
\begin{minipage}{9cm}
\epsfxsize=3.3in
\epsffile{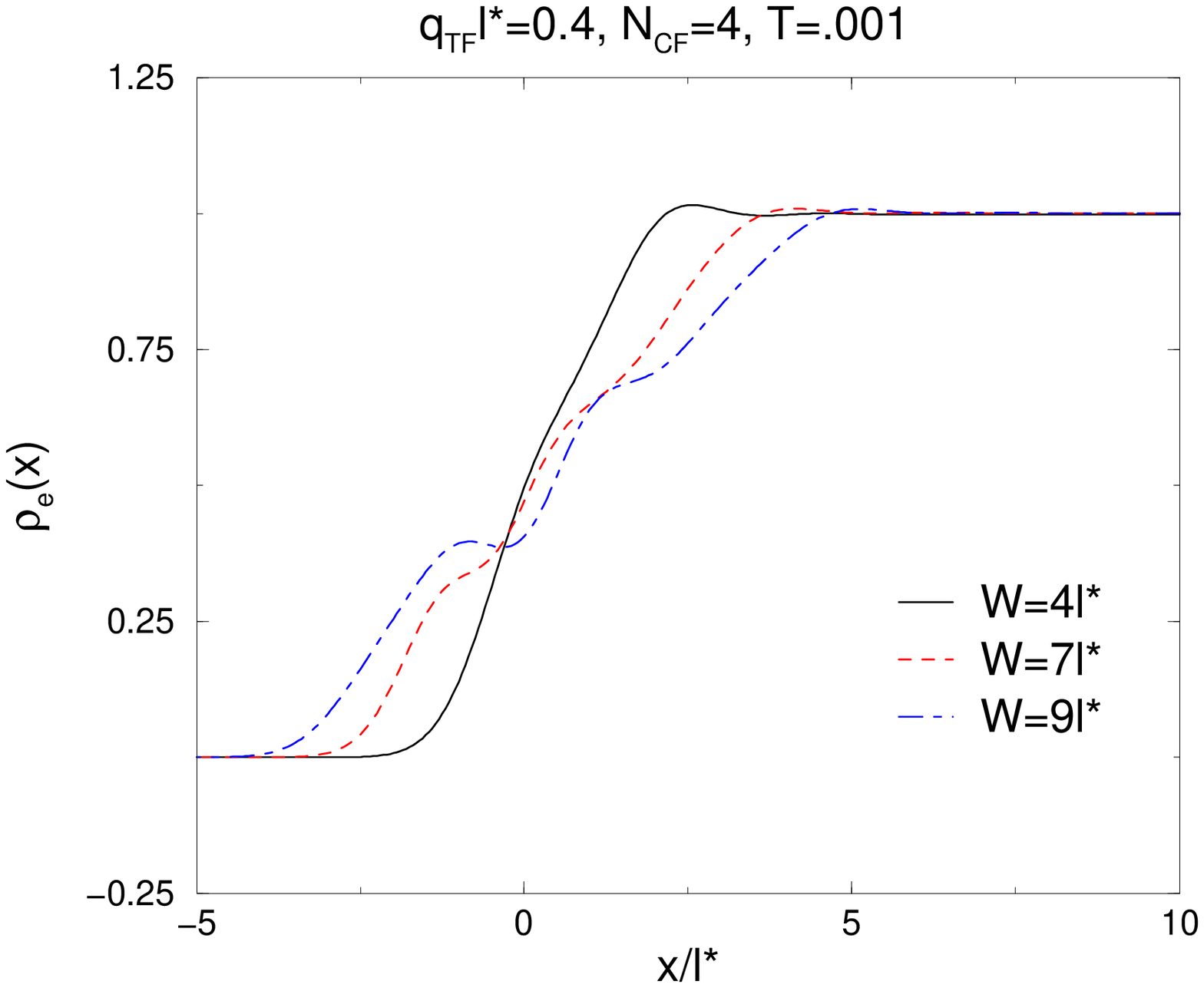}
\end{minipage}
\end{minipage}
\caption{Edge-reconstruction for $\nu=1/3$ with screened Coulomb interaction: The figure on left (right) 
shows the CF occupation number (electron density) for various widths $W$ of the background charge-density. 
The temperature is measured in the units of electron-electron interaction and is much smaller than the 
bulk-gap for the CF Landau levels.~\cite{conserv} For $W\geq 6l^{*}$ the edge reconstructs and a strip of 
charge with width $l^{*}$ moves away a distance $d\approx 2l^{*}$ from the bulk.}
\label{fig: nu1thomas}
\end{center}
\end{figure}

\begin{figure}[t]
\begin{center}
\begin{minipage}{20cm}
\begin{minipage}{9cm}
\epsfxsize=3.3in
\epsffile{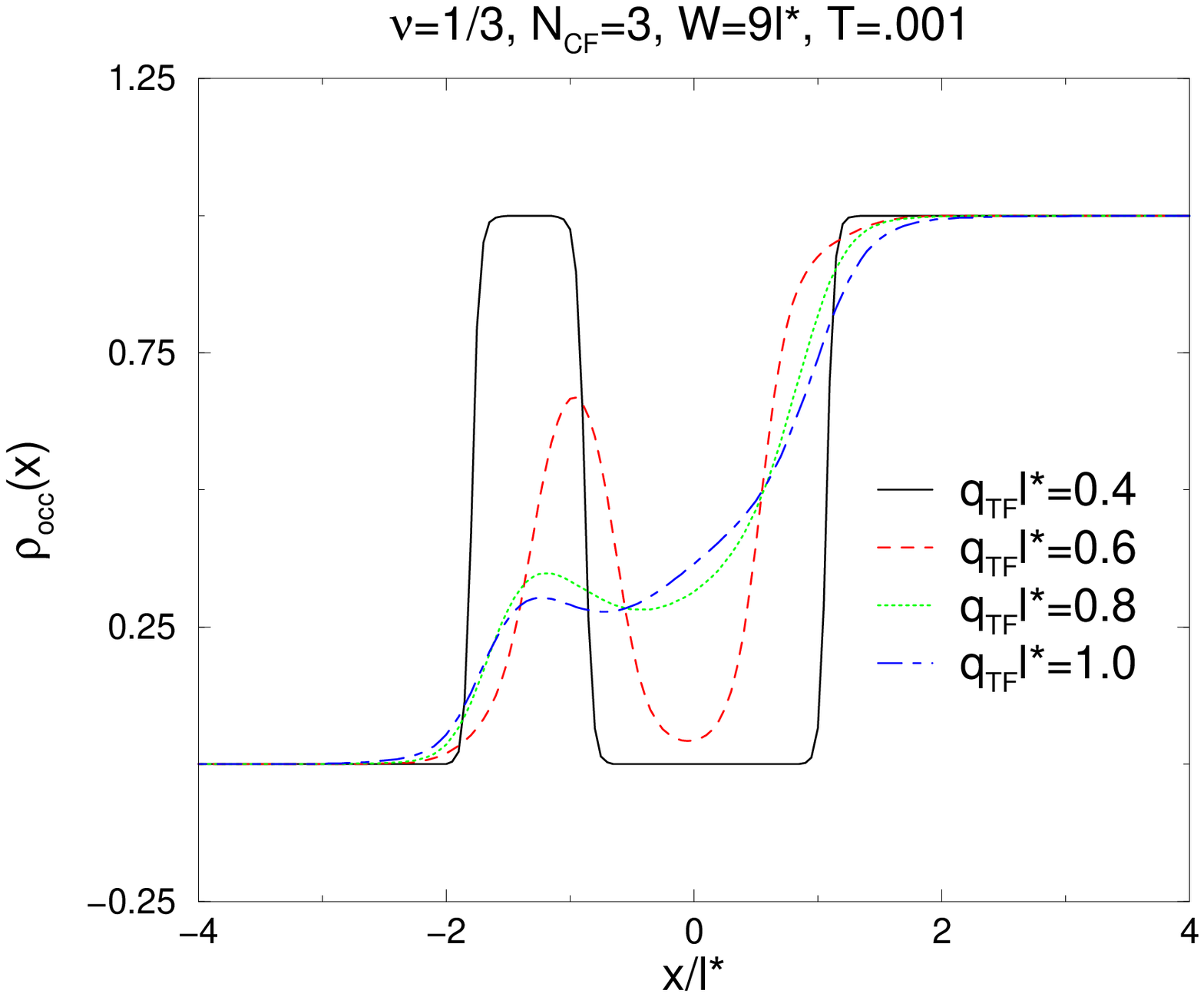}
\end{minipage}
\begin{minipage}{9cm}
\epsfxsize=3.3in
\epsffile{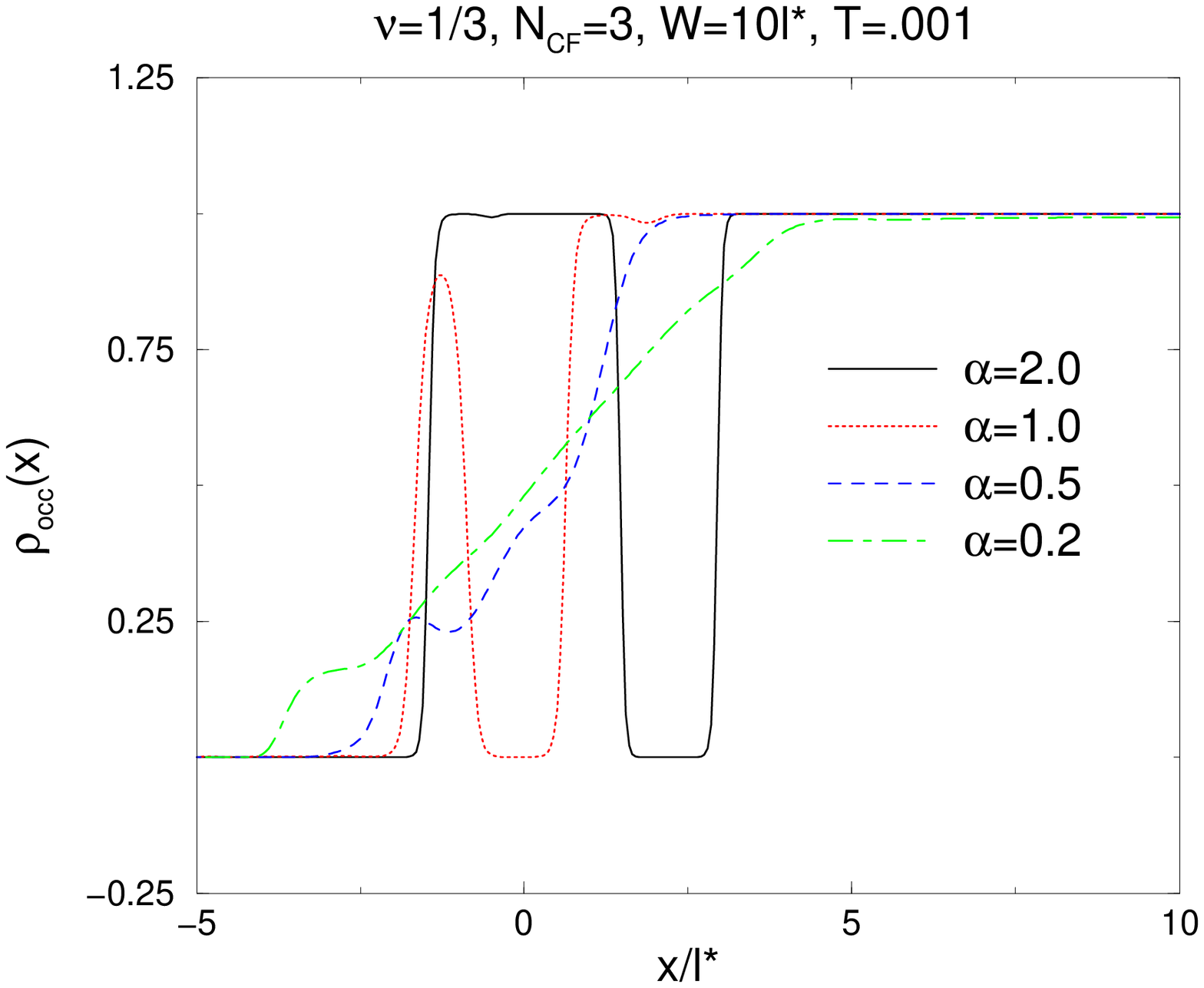}
\end{minipage}
\end{minipage}
\caption{Range dependence of the edge-reconstruction at $\nu=1/3$: The left (right) figure shows the CF 
occupation number for screened Coulomb (Gaussian) interactions. In each case, we see that the 
edge-reconstruction becomes pronounced with increasing range. Stated differently, the critical width 
$W^{*}$ at which reconstruction takes place increases with decreasing range of the interaction.}
\label{fig: range}
\end{center}
\end{figure}

\begin{figure}[t]
\begin{center}
\begin{minipage}{20cm}
\begin{minipage}{9cm}
\epsfxsize=3.3in
\epsffile{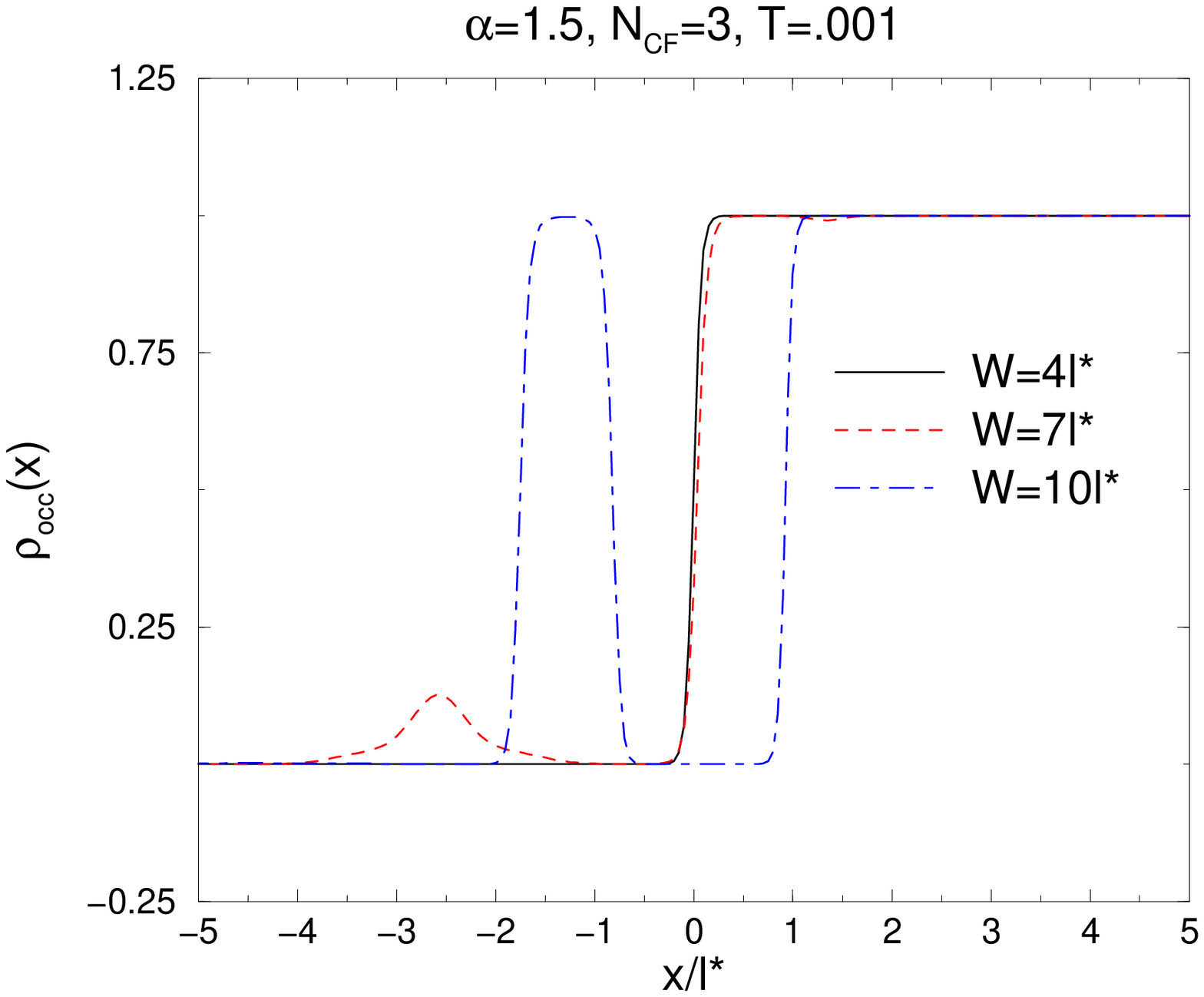}
\end{minipage}
\begin{minipage}{9cm}
\epsfxsize=3.3in
\epsffile{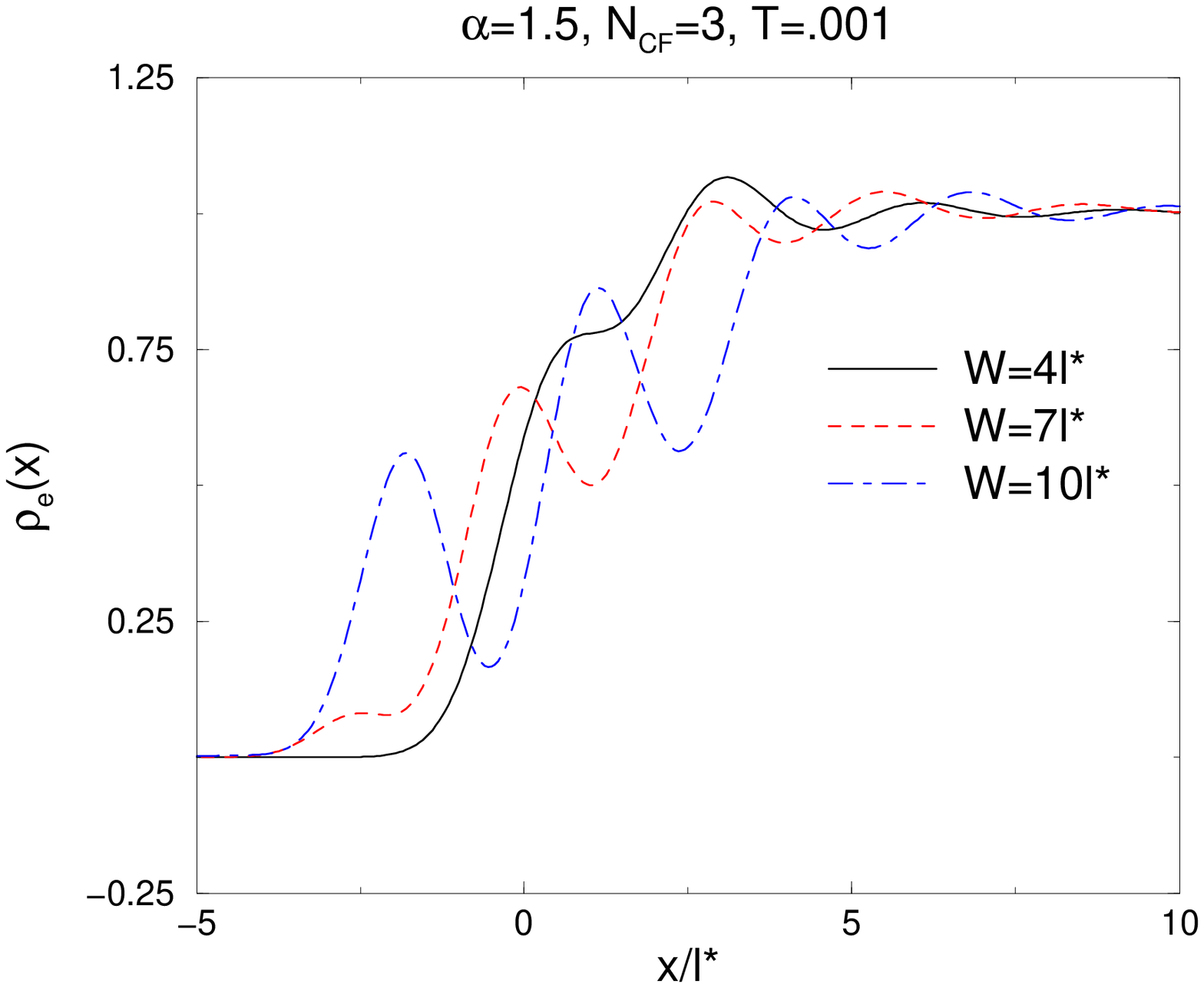}
\end{minipage}
\end{minipage}
\caption{Edge-reconstruction for $\nu=1/3$ with Gaussian interaction: The figure on left (right) shows 
the CF occupation number (electron density) for various widths $W$. The edge reconstructs for 
$W\geq 7l^{*}$. This edge reconstruction is reflected in the non-monotonic behavior of the electron density. 
The marked difference between density profiles for the Gaussian and the screened Coulomb interactions is 
due to the difference between the self-consistent density matrices in respective cases, even though the 
resultant CF occupation-number profiles are similar.}
\label{fig: nu1gauss}

\end{center}
\end{figure}

\begin{figure}[t]
\begin{center}
\begin{minipage}{20cm}
\begin{minipage}{9cm}
\epsfxsize=3.3in
\epsffile{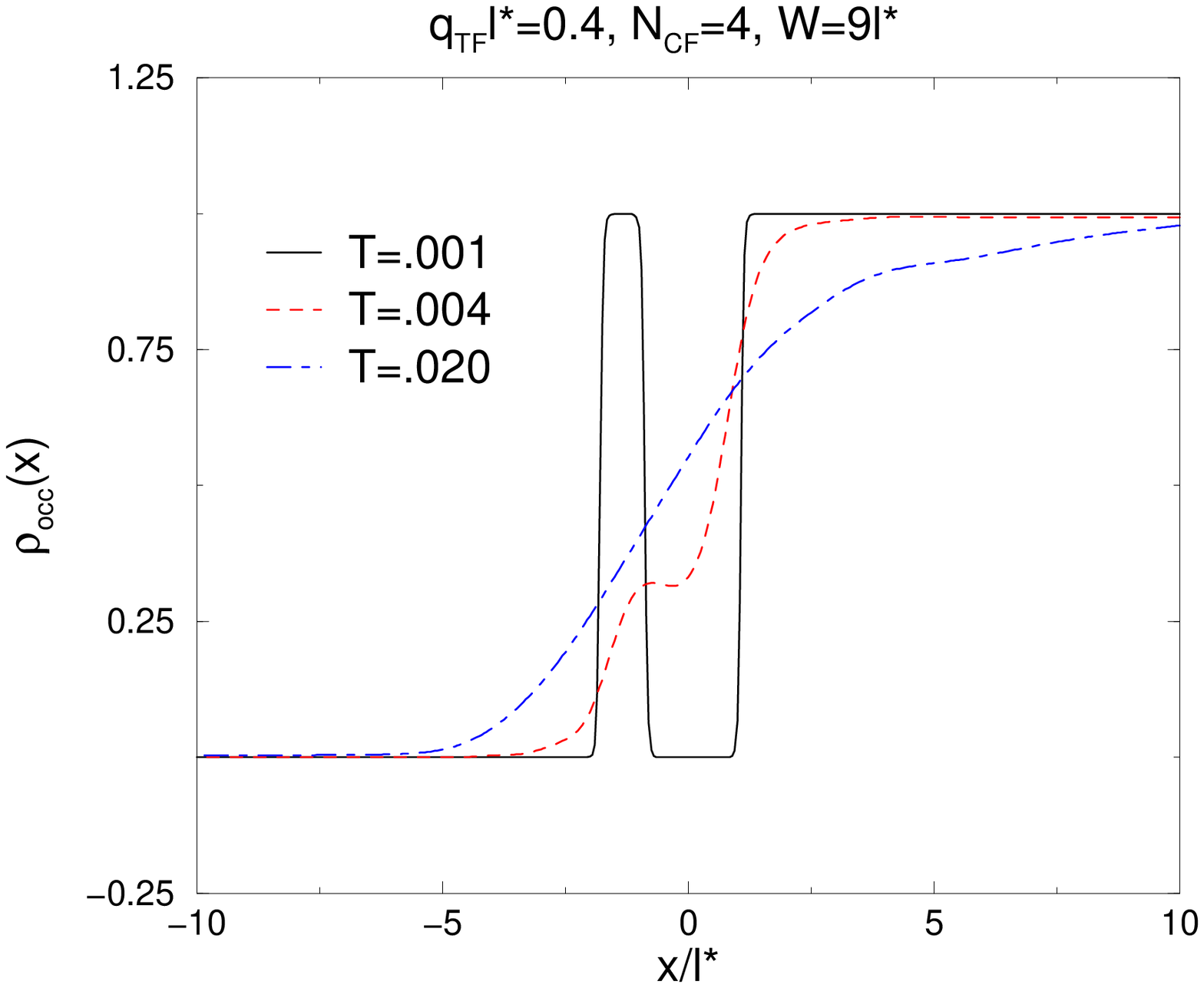}
\end{minipage}
\begin{minipage}{9cm}
\epsfxsize=3.3in
\epsffile{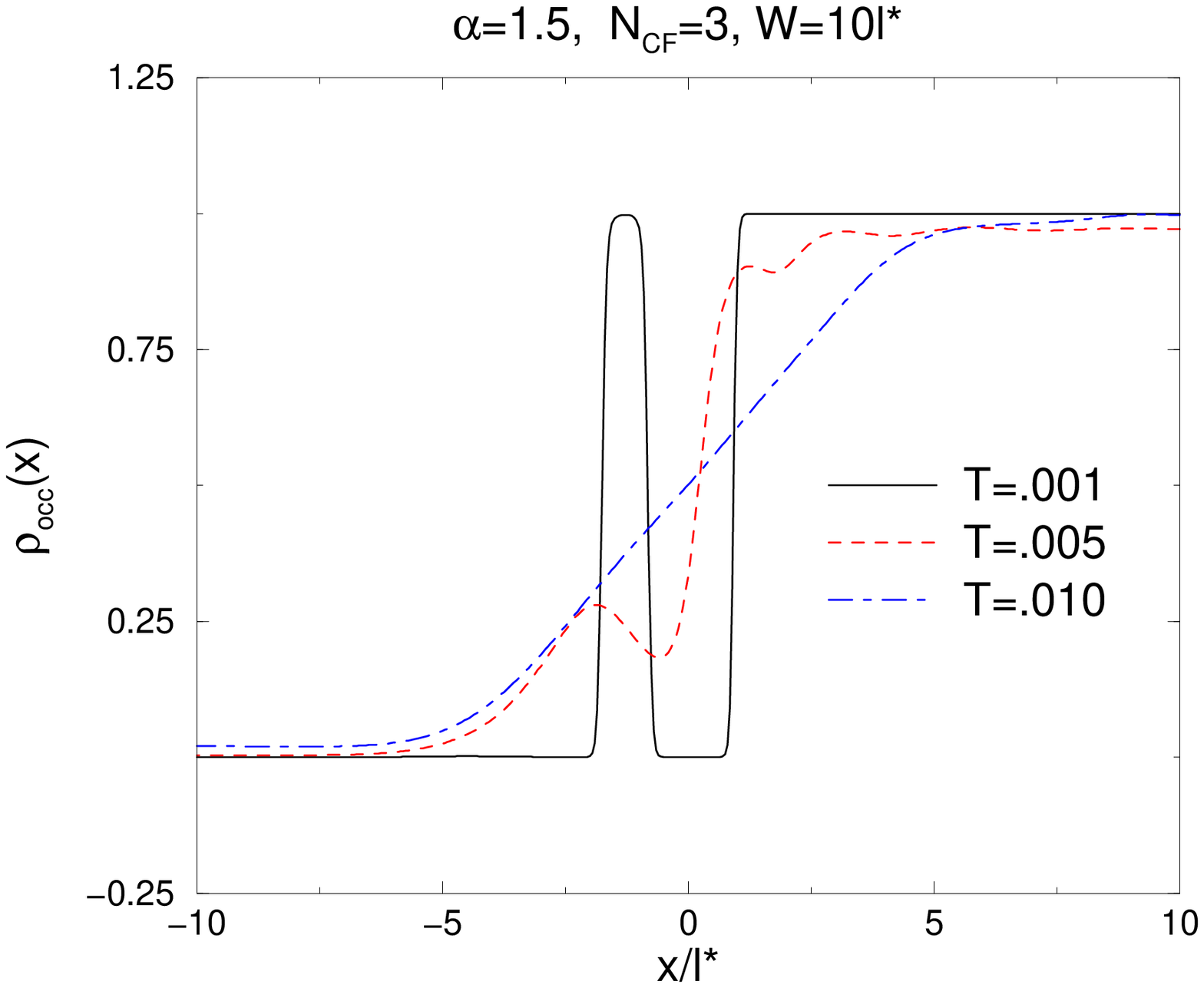}
\end{minipage}
\end{minipage}
\caption{Temperature dependence of the edge-reconstruction at $\nu=1/3$: The left (right) figure shows the 
CF occupation numbers with increasing temperature for screened Coulomb (Gaussian) interactions. In both 
cases, we see that the edge-reconstruction disappears with increasing temperature.}
\label{fig: temp}
\end{center}
\end{figure}

Figure~\ref{fig: nu1thomas} shows the CF occupation number and the electron density for a system with 
$q_{TF}l^{*}=0.4$ and with four CF Landau levels, $N_{CF}=4$. As we can see, the edge undergoes 
reconstruction which is manifest in a strip of charge moving away from the bulk. A systematic study with 
increasing number of CF Landau levels included in the Hartree-Fock iterations shows that the reconstruction 
is stable with increasing $N_{CF}$. The left panel in Fig.~\ref{fig: range} shows the dependence of the 
reconstruction on the Thomas-Fermi cutoff $q_{TF}$. It is clear that as $q_{TF}\rightarrow 0$, the 
reconstruction becomes easier and more pronounced; however, it is present even at $q_{TF}l^{*}=1$. We find 
that the edge reconstruction is also obtained with a Gaussian interaction (Fig.~\ref{fig: nu1gauss}). As is 
the case with the screened Coulomb interaction, the edge-reconstruction is facilitated by an increasing range 
(Right panel in Fig.~\ref{fig: range}). Note that as 
$\alpha\rightarrow 0$ the electron-electron interaction becomes local in real-space and the reconstruction 
disappears. The typical temperature dependence of the edge-reconstruction for both types of interactions is 
shown in Fig.~\ref{fig: temp}. In each case, the edge-reconstruction disappears with increasing temperature. 
This is expected since the system becomes more and more classical with increasing temperature. We also verify 
that when the temperature is reduced, the self-consistent CF occupation numbers tend towards 0 or 1; however 
the qualitative evolution of edge-reconstruction with varying range of interaction, $q_{TF}$ or $\alpha$, 
remains unchanged down to $T=0$.

\begin{figure}[t]
\begin{center}
\begin{minipage}{20cm}
\begin{minipage}{9cm}
\epsfxsize=3.3in
\epsffile{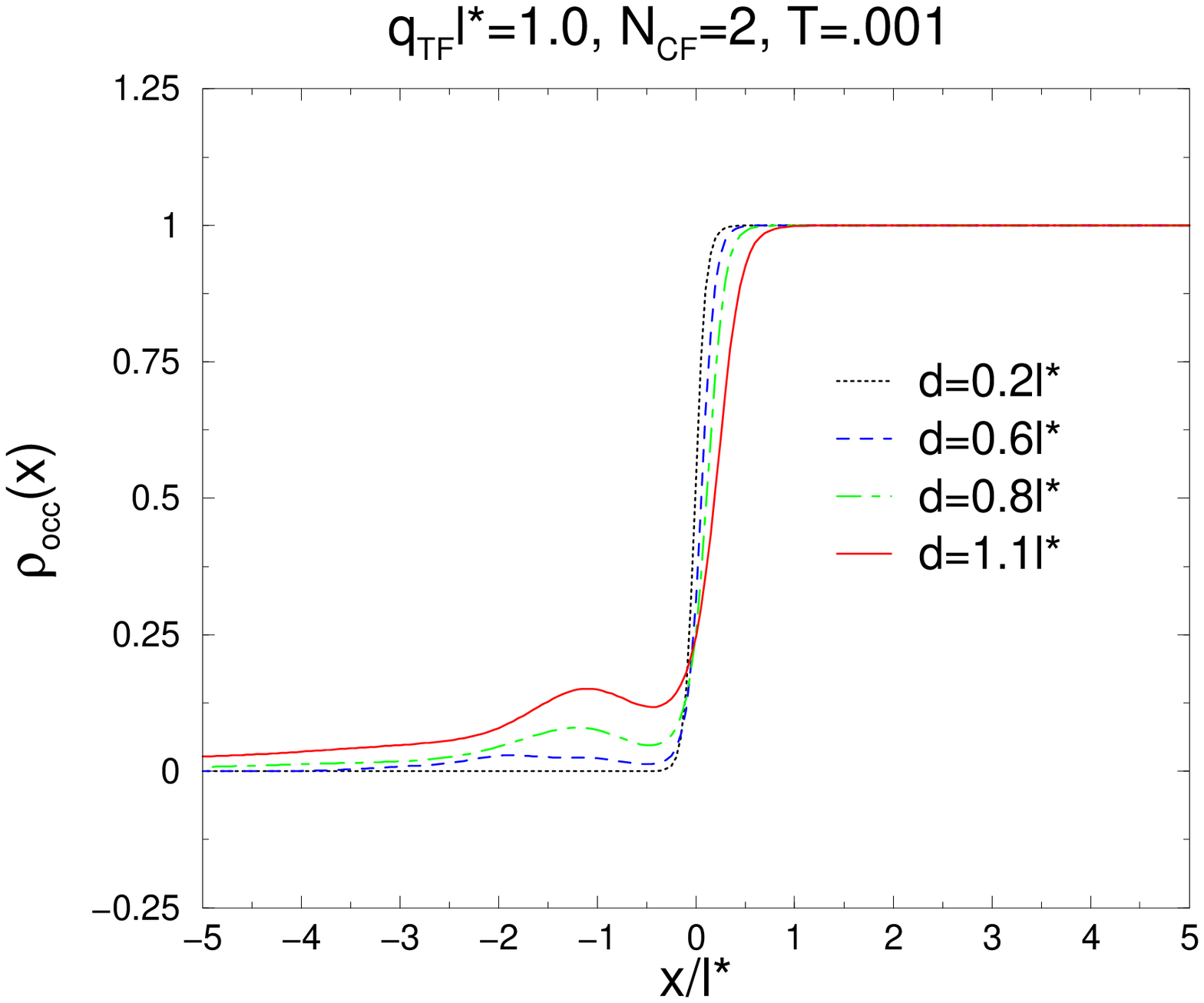}
\end{minipage}
\begin{minipage}{9cm}
\epsfxsize=3.3in
\epsffile{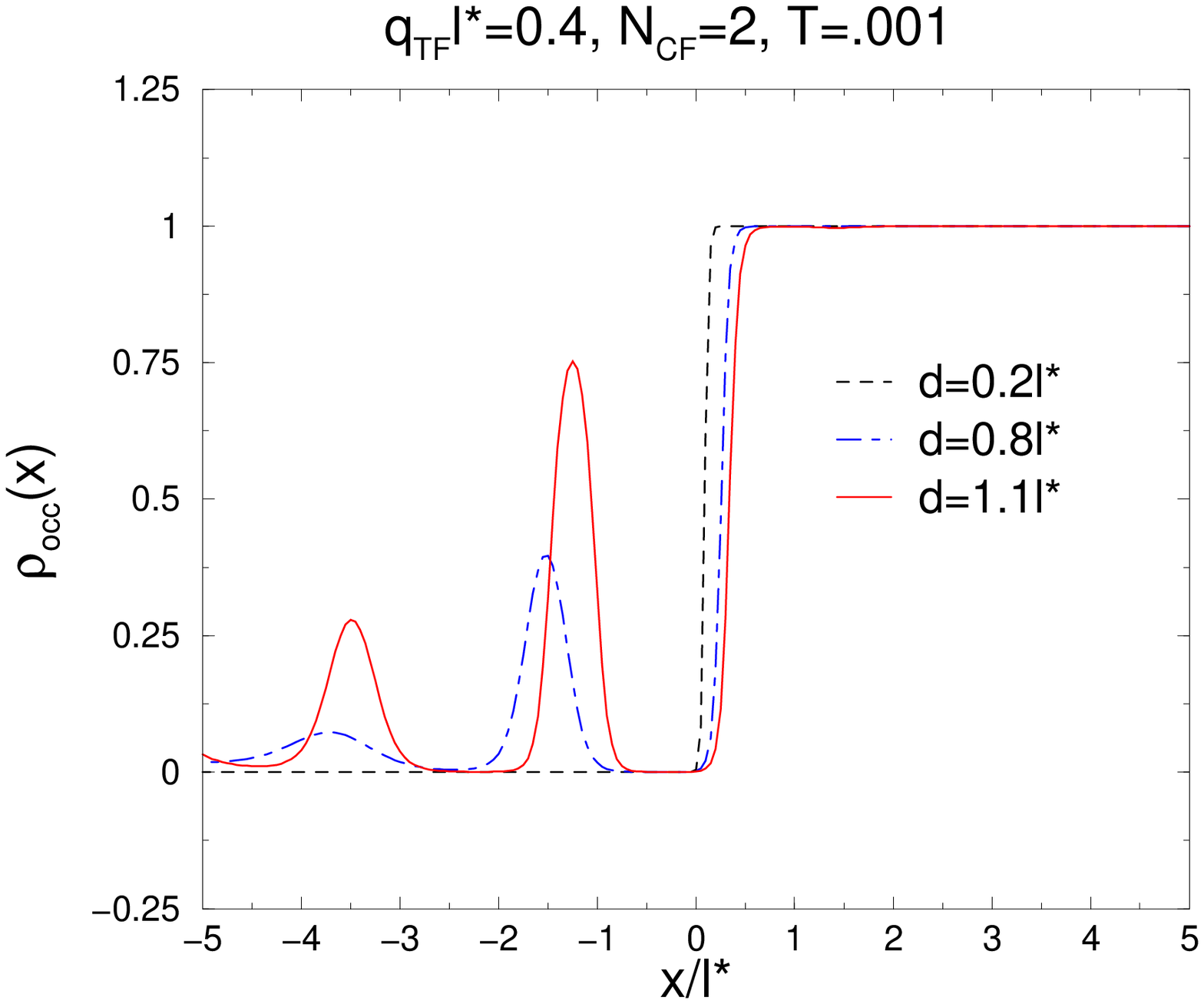}
\end{minipage}
\end{minipage}
\caption{Edge-reconstruction for $\nu=1/3$ with the background-layer distance $d$ away. The figure on 
left (right) shows the CF occupation numbers for screened Coulomb interaction with $q_{TF}l^*=1.0$ 
($q_{TF}l^*=0.4$). The edge undergoes reconstruction at a critical distance $d_{cr}(q_{TF})$ which 
decreases as $q_{TF}l^*\rightarrow 0$. These results are consistent with those obtained by Wan 
{\it et al.}~\cite{wan}; however, they also show that reconstruction does not require a 
{\it pure} Coulomb interaction. The multiple reconstructions for $q_{TF}l^*=0.4$ give way to a single 
strip of charge separated from the bulk when $N_{CF}$ is increased and temperature is decreased. }
\label{fig: displaced}
\end{center}
\end{figure}

Next we compare the edge-reconstruction results obtained using this approach with those obtained by Wan 
{\it et al.} using exact diagonalization.~\cite{wan,wan2} To this end, we assume that the background 
charge-density is offset by a distance $d$ from the 2D electron gas. Thus, we use 
$V_{eb}(\vec{q})=V_{ee}(\vec{q})e^{-qd}$ and study the edge profile as a function of increasing $d$. 
Figure~\ref{fig: displaced} shows the CF occupation numbers for $q_{TF}l^{*}=1.0$ (left) and $q_{TF}l^{*}=0.4$ 
(right). We see that the $\nu=1/3$ edge undergoes reconstruction as $d$ increases, and that the critical 
value $d(q_{TF})$ decreases with decreasing $q_{TF}$. These results, in the limit $q_{TF}l^*\rightarrow 0$, 
are quantitatively consistent with those obtained by Wan {\it et al.}~\cite{wan,wan2} and show that 
the edge reconstructs even with a screened Coulomb interaction. As mentioned in the introduction, the 
edge-reconstruction is an inherently quantum phenomenon; pure electrostatic considerations require, at least 
for long-ranged interactions, that the electron-density profile follow the background-charge profile. The 
energy cost of deviations from the background charge-density is compensated by the exchange energy gained by 
the electrons because of the step-like profile. {\it These results show that edge reconstructions at 
$\nu=1/3$ are generic and that the long-range nature of the electron-electron interactions is not 
instrumental to their existence.} 

\begin{figure}[t]
\begin{center}
\begin{minipage}{20cm}
\begin{minipage}{9cm}
\epsfxsize=3.3in
\epsffile{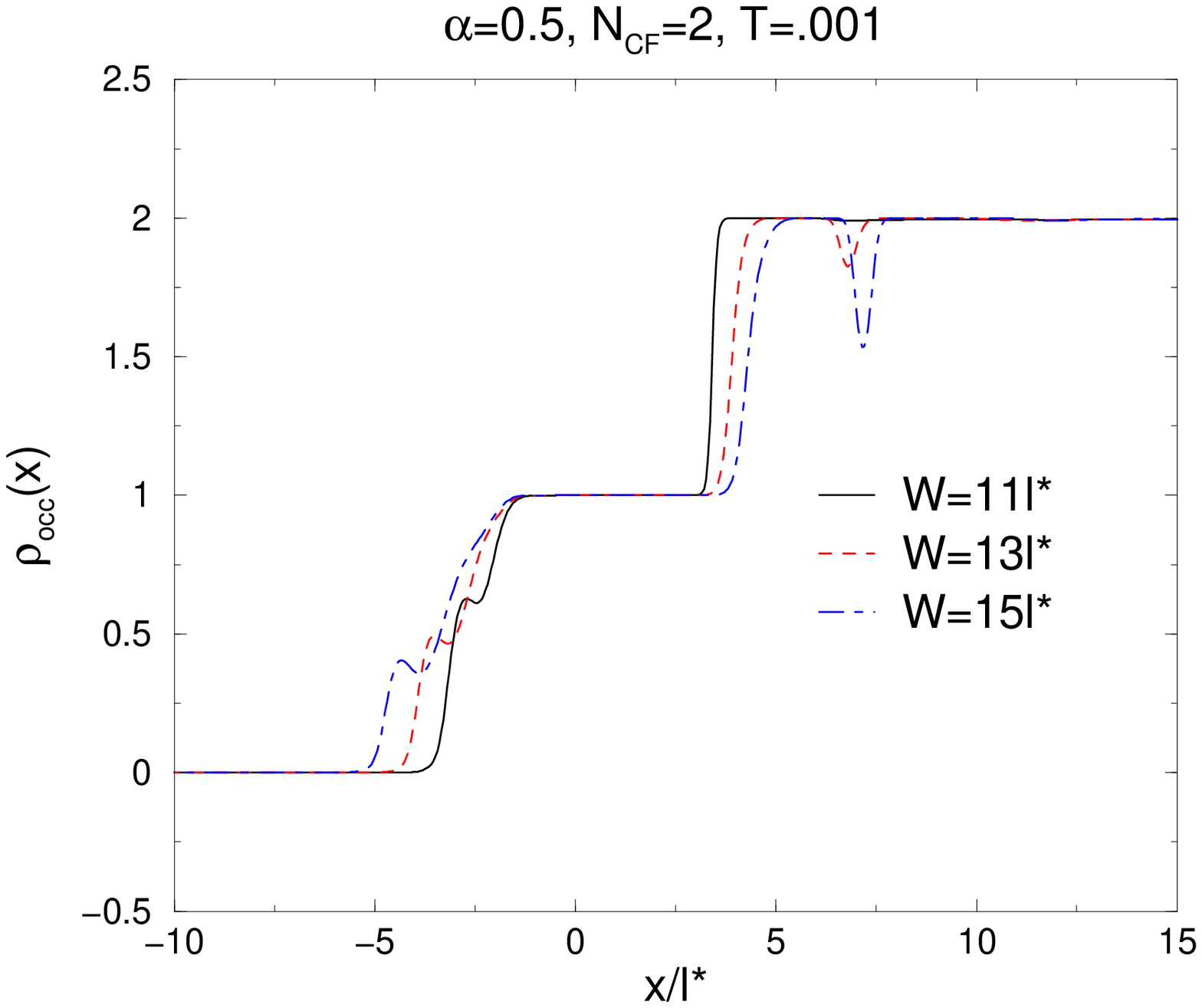}
\end{minipage}
\begin{minipage}{9cm}
\epsfxsize=3.3in
\epsffile{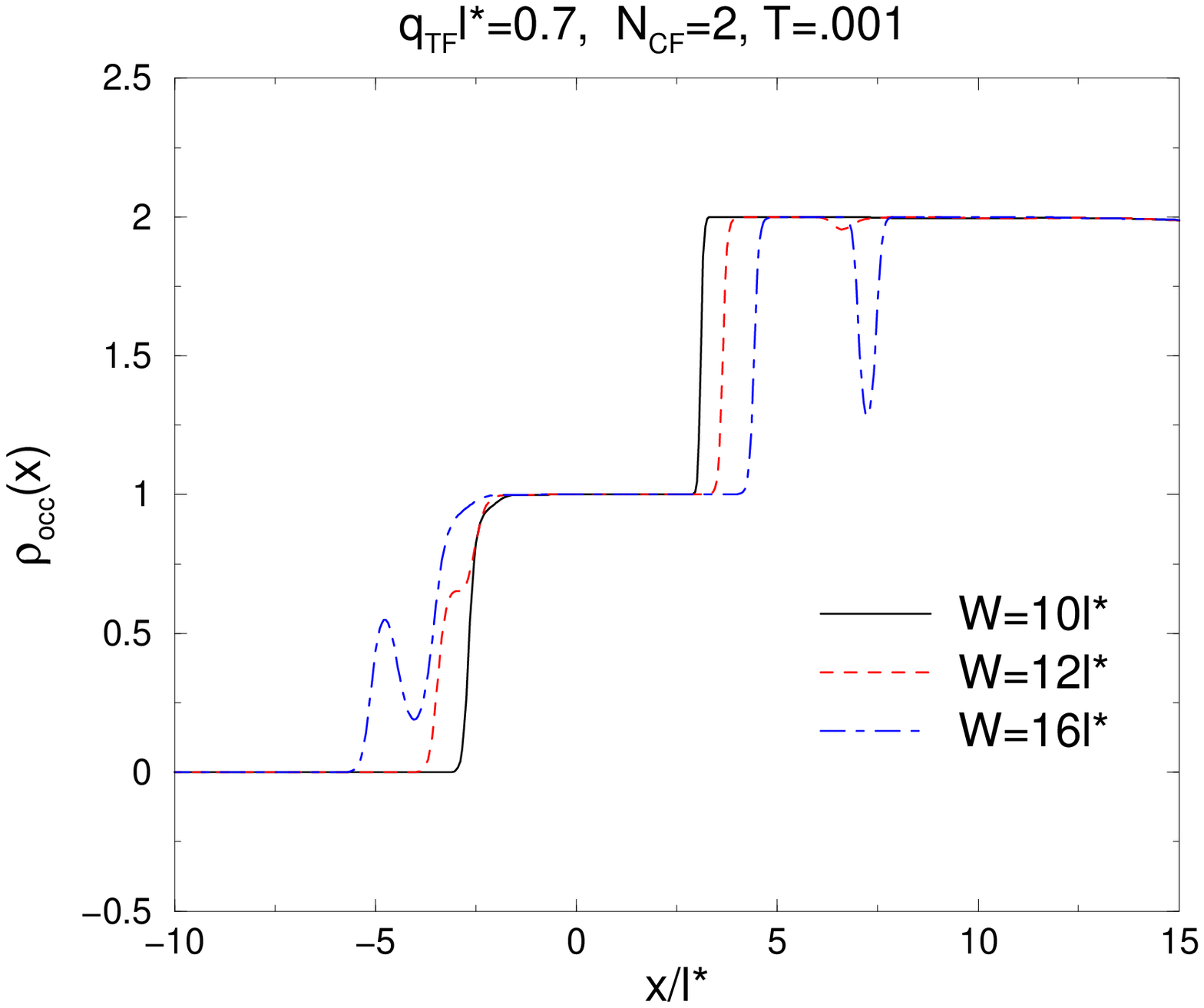}
\end{minipage}
\end{minipage}
\begin{minipage}{20cm}
\begin{minipage}{9cm}
\epsfxsize=3.3in
\epsffile{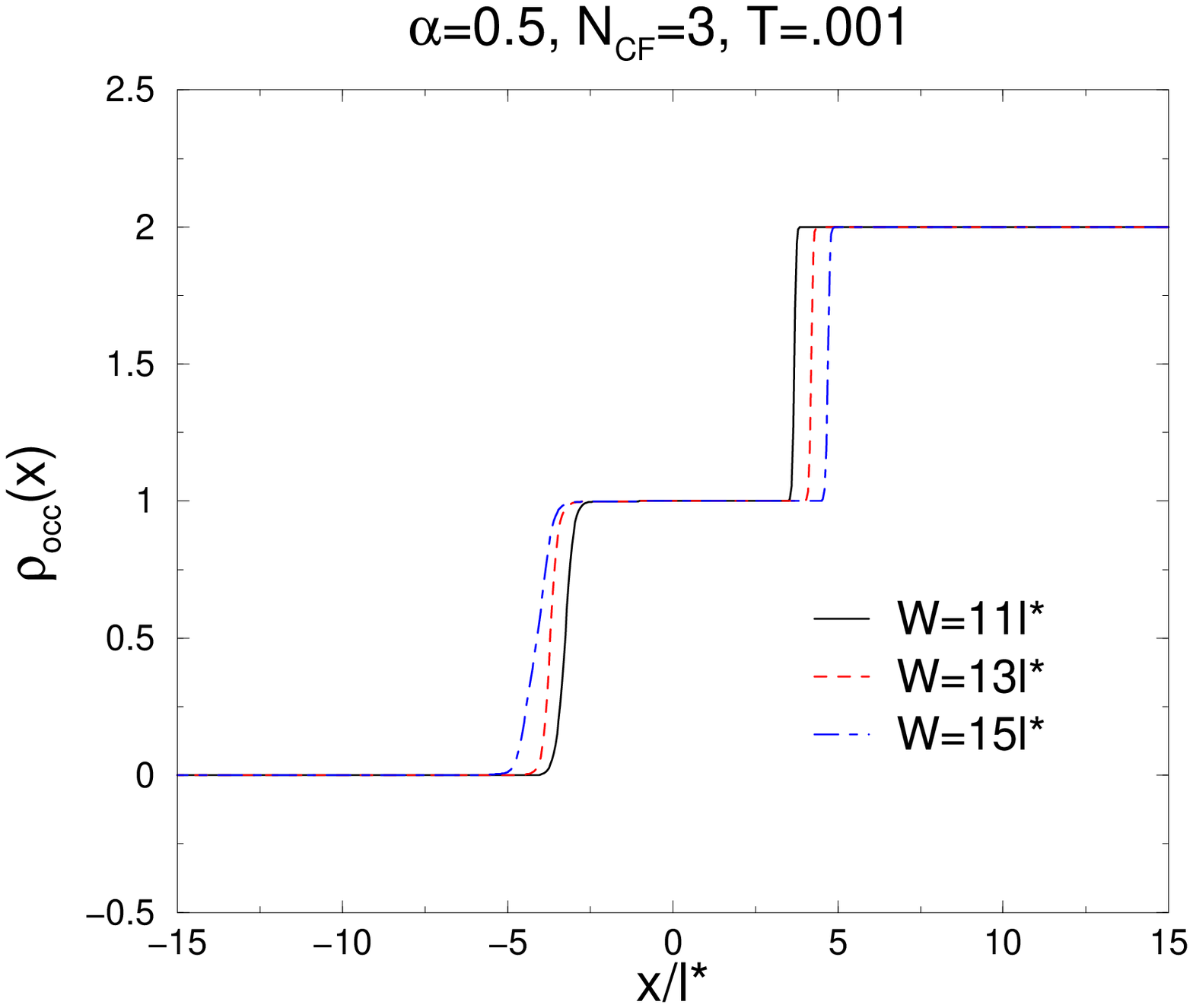}
\end{minipage}
\begin{minipage}{9cm}
\epsfxsize=3.3in
\epsffile{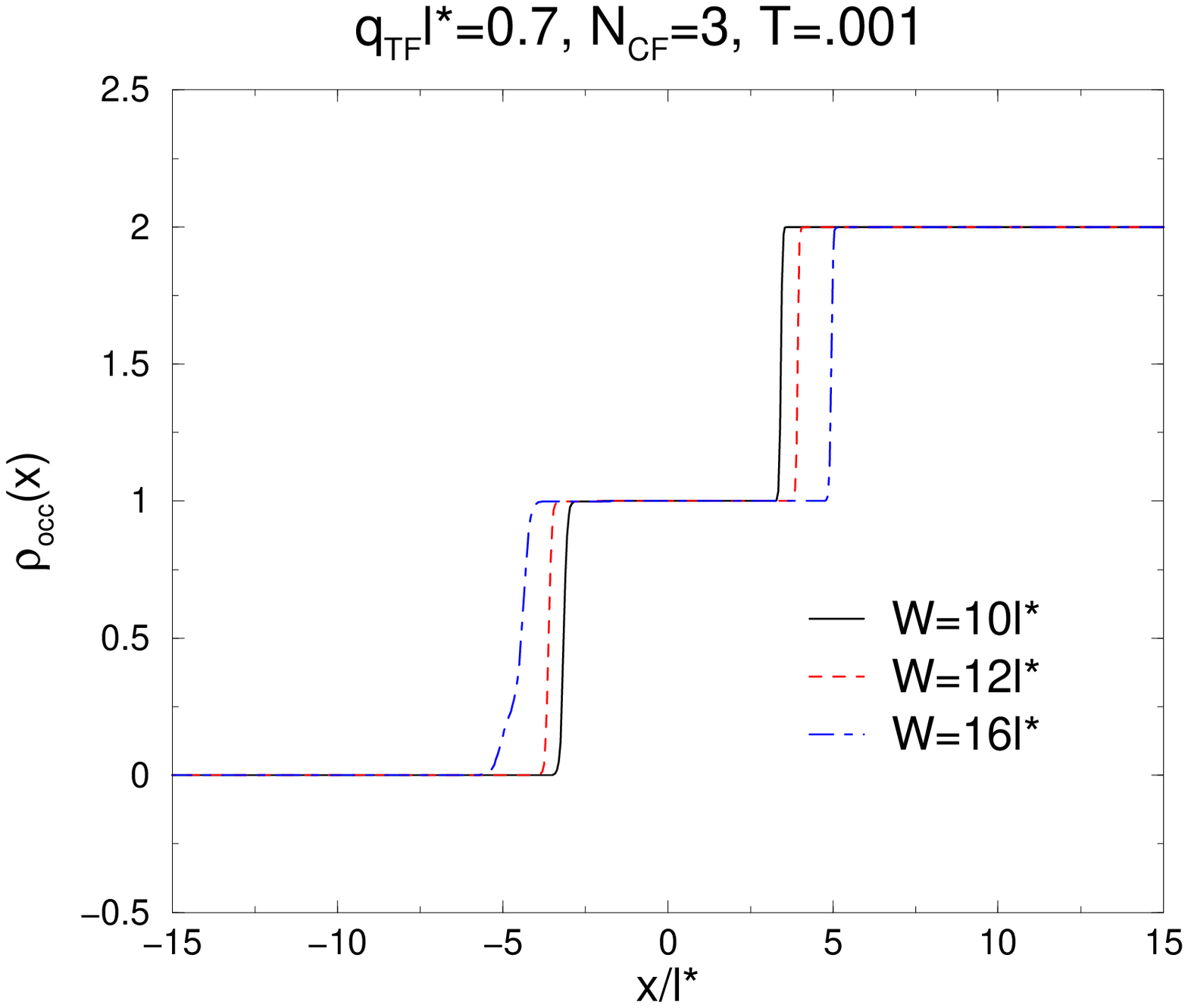}
\end{minipage}
\end{minipage}
\caption{Edge-states for the $\nu=2/5$: We find that with $N_{CF}=2$ (top panel), Gaussian (left) and 
Thomas-Fermi (right), interactions lead to edge reconstructions. However, as the bottom panel shows, these 
reconstructions are {\it not} stable with increasing $N_{CF}$, and therefore should be treated as the 
artifact of the approximation. Calculations with higher CF Landau levels up to $N_{CF}=5$ show that the 
edge profile for $\nu=2/5$ remains smooth.}
\label{fig: nu2}
\end{center}
\end{figure}

\begin{figure}[t]
\begin{center}
\begin{minipage}{20cm}
\begin{minipage}{9cm}
\epsfxsize=3.3in
\epsffile{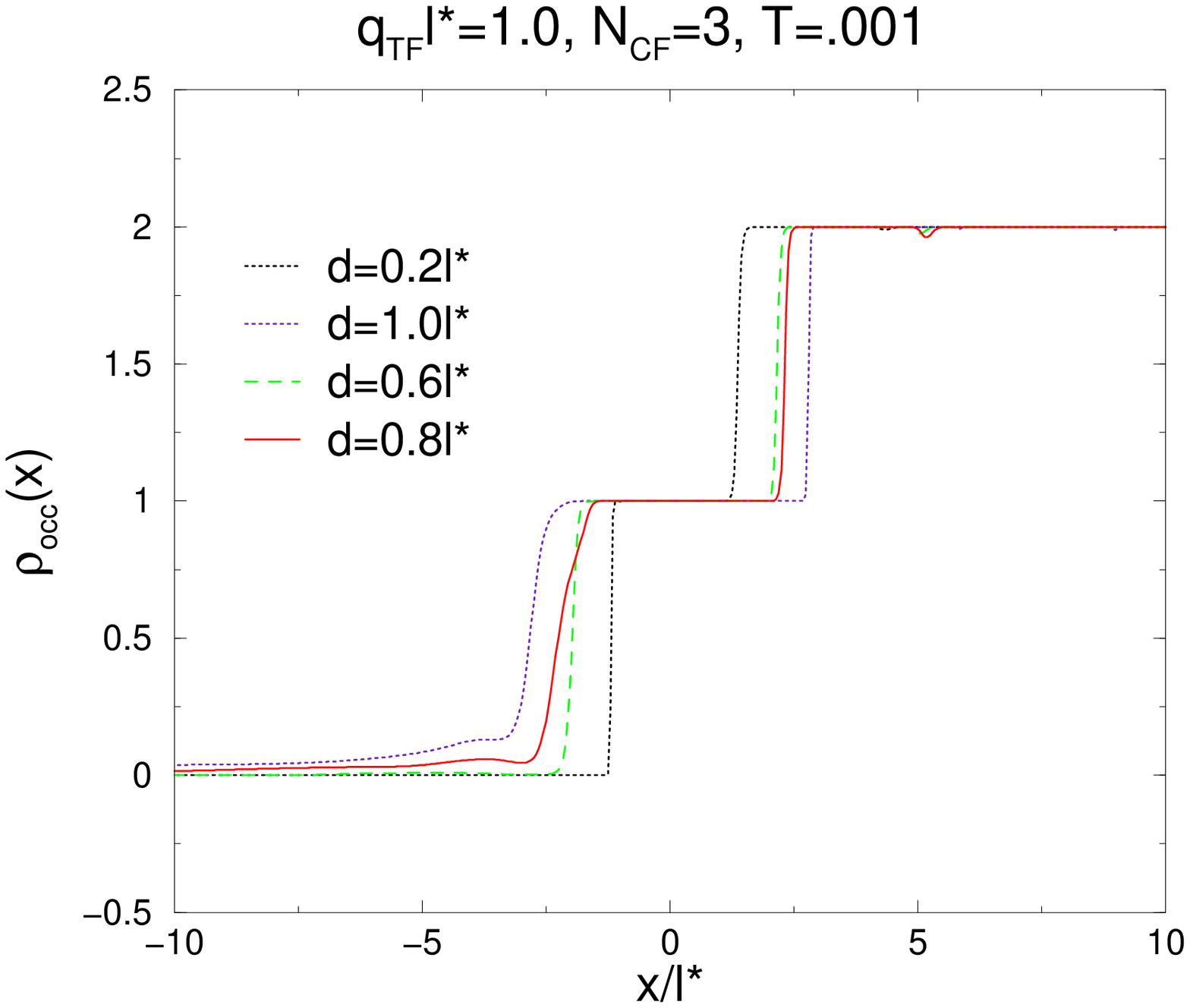}
\end{minipage}
\begin{minipage}{9cm}
\epsfxsize=3.3in
\epsffile{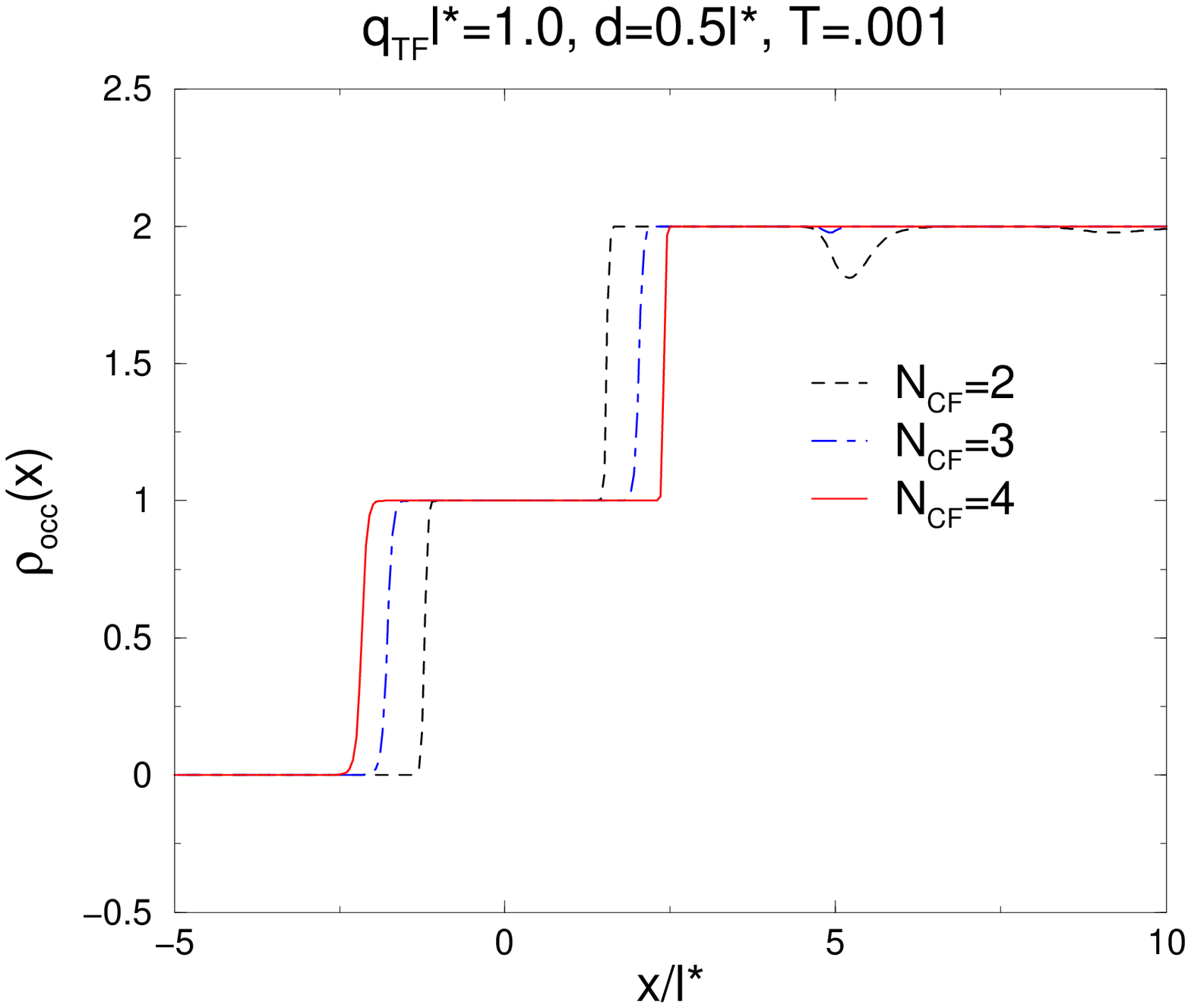}
\end{minipage}
\end{minipage}
\caption{Edge-states for the $\nu=2/5$ with a displaced background: The left figure shows the CF occupation 
number for various $d$ with $N_{CF}=3$. We find that the region of unit CF occupancy increases with 
increasing $d$, as in the case of Fig.~\ref{fig: nu2}. The realistic sample values $d\sim 5l^*$ are very 
difficult to access due to numerical limitations. The figure on the right shows that the reconstruction 
appearing near $x\approx 6l^*$ is indeed spurious, and is not stable with increasing $N_{CF}$. Therefore the 
edge profile for $\nu=2/5$ remains smooth with increasing $d$.}
\label{fig: nu2d}
\end{center}
\end{figure}

\begin{figure}[t]
\begin{center}
\begin{minipage}{20cm}
\begin{minipage}{9cm}
\epsfxsize=3.3in
\epsffile{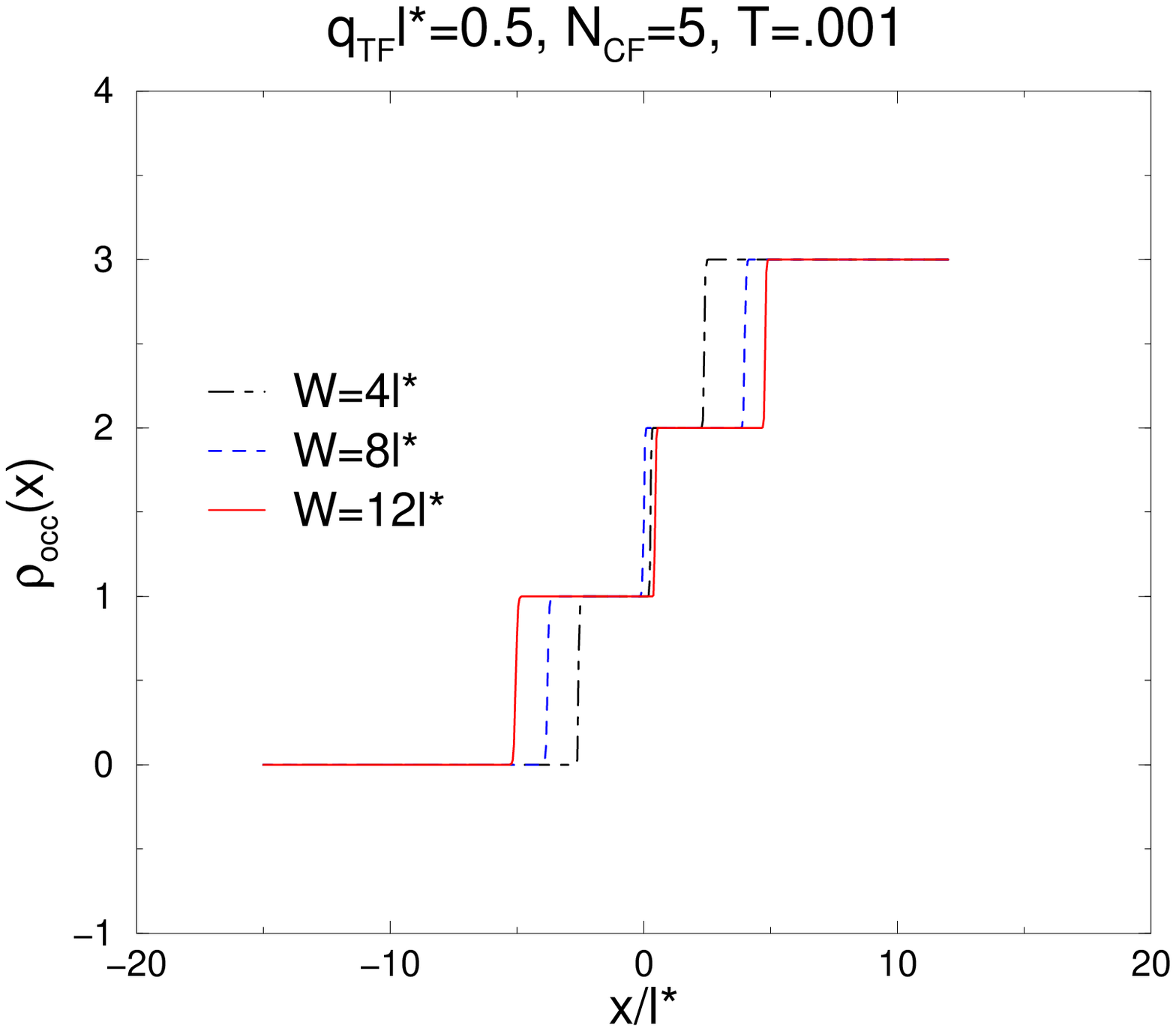}
\end{minipage}
\begin{minipage}{9cm}
\epsfxsize=3.3in
\epsffile{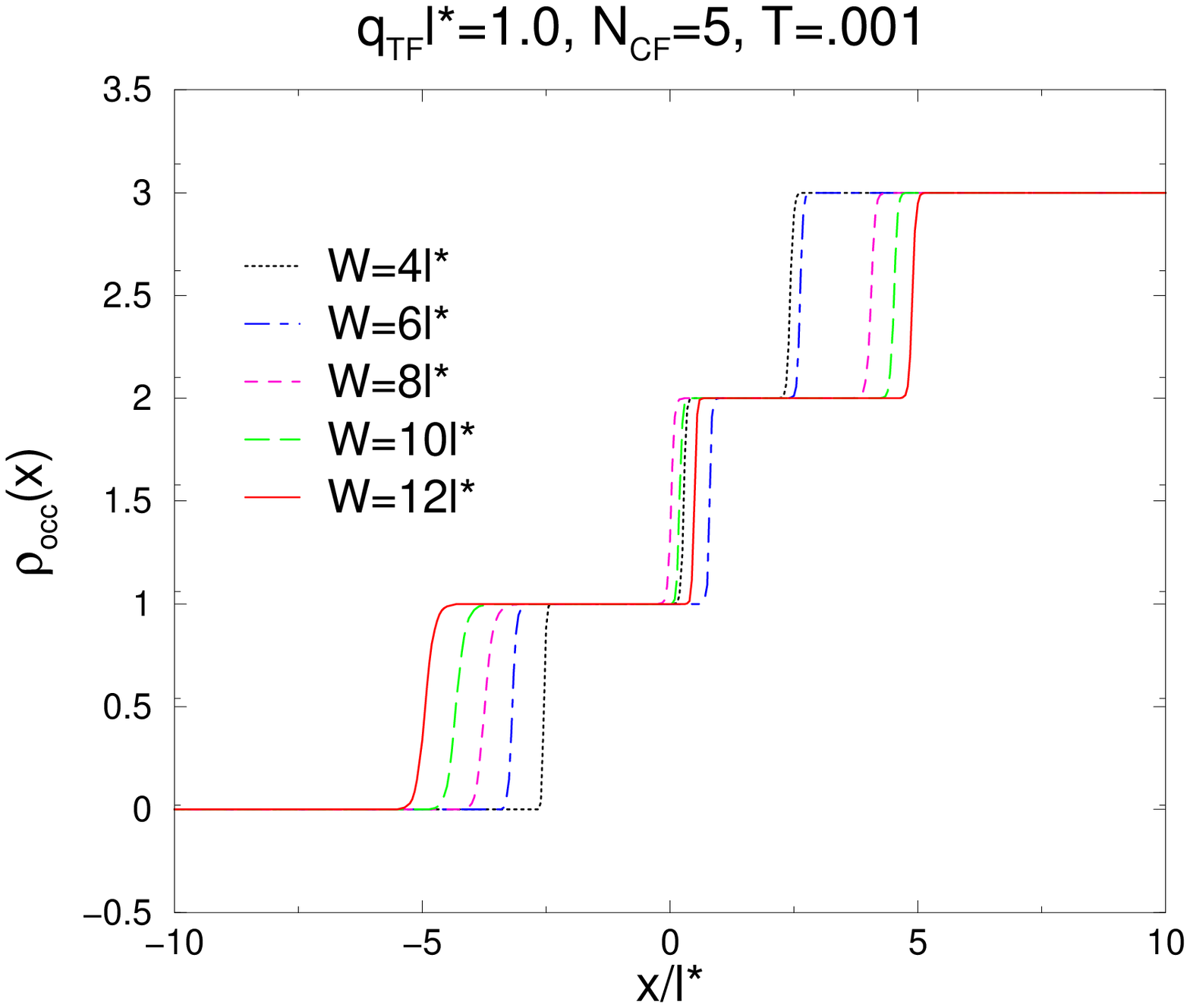}
\end{minipage}
\end{minipage}
\caption{Edge-states for a spin-polarized $\nu=3/7$: The left (right) figure shows the CF occupation numbers 
for a screened Coulomb interaction with $q_{TF}l^*=0.5$ ($q_{TF}l^*=1.0$) for $N_{CF}=5$. It is clear from 
both figures that the system is robust against reconstruction. However, as the figure on the right shows, 
there is a critical value of $W$ ($\approx 8l^*$) when the system, {\it instead of undergoing reconstruction}, 
widens the regions having $\rho_{occ}(x)=1$ and $\rho_{occ}(x)=2$ and maintains a smooth edge profile.}
\label{fig: nu3}
\end{center}
\end{figure}

Now we consider the edge states in a quantum Hall system at higher fractional filling factors, $\nu=2/5$ 
and $\nu=3/7$. These systems are mapped onto composite fermion systems at filling factor $p=2$ and $p=3$ 
respectively. In this paper, we assume that all electron spins are fully polarized due to the strong external 
magnetic field, and therefore, in the ground state, the first two (three) CF Landau levels are occupied. 
The top panel in Fig.~\ref{fig: nu2} shows the self-consistent CF occupations, with $N_{CF}=2$, for Gaussian 
interaction with range $\alpha=0.5$ (left) and for Thomas-Fermi interaction with cutoff 
$q_{TF}l^{*}=0.7$ (right). It appears from these figures that the edge reconstructs with increasing $W$ for 
different values of $\alpha$ and $q_{TF}$. This, however, is an artifact of the approximation. As the 
bottom panel in Fig.~\ref{fig: nu2} shows, when we include the higher CF Landau levels, the spurious edge 
reconstruction disappears. Instead, the effect of increasing $W$ is manifest through the increasing width of 
the region where CF occupation number is unity, $\rho_{CF}(x)=1$. The same physics appears when we consider 
a system with displaced background layer (Fig.~\ref{fig: nu2d}). The figure on the left shows that with 
increasing background-layer distance $d$, the region where $\rho_{CF}=1$ widens. The figure on the right 
illustrates how the spurious edge-reconstruction is suppressed with increasing $N_{CF}$, the number of Landau 
levels used in the Hartree-Fock iterations. In experimental samples, the background layer is situated at a 
distance $d\sim 5l^*$ away from the 2D electron gas. It is very difficult to access this regime in the 
Hartree-Fock calculations due to numerical limitations. However, these results suggest that the spin-polarized 
$\nu=2/5$ state is robust against edge reconstruction. We obtain similar 
results of the spin-polarized $\nu=3/7$ edge. Figure~\ref{fig: nu3} shows the CF occupations numbers for a 
screened Coulomb interaction with $q_{TF}l^*=0.5$ (left) and $q_{TF}l^*=1.0$ (right) for $N_{CF}=5$. It is 
obvious from the CF occupation numbers that the edge remains smooth with increasing $W$; however if we reduce 
the number of CF Landau levels used in the iterations to $N_{CF}=3$, the spurious edge reconstructions (similar 
to those seen in Fig.~\ref{fig: nu2} top panel) reappear in this case as well. The results for $q_{TF}l^*=1.0$ 
(right) show that there is a critical $W$ ($\sim 8l^*$) when the system, instead of undergoing reconstruction, 
suddenly widens the regions where $\rho_{CF}(x)=1$ or $\rho_{CF}(x)=2$. This freedom of varying $\rho_{CF}(x)$ 
in steps 
of one to gain exchange energy, without undergoing reconstruction, is available only at higher filling factors 
where more than one CF Landau levels are occupied in the bulk. {\it Thus, we find that spin-polarized 
quantum Hall liquids at $\nu=2/5$ and $\nu=3/7$ are more robust against reconstruction.}

\section{Remarks}
\label{sec: remarks}
In this paper, we have presented a microscopic analytical calculation of the edge states in the fractional 
quantum Hall regime using the extended Hamiltonian theory.~\cite{hamreview} This approach is particularly 
suited to the study of edge states at different filling factors and with different system parameters. 
Although we have only considered systems where the neutralizing background provides the confining barrier at 
the edge, It is easy to incorporate an external barrier in this formalism and it does not change the results 
qualitatively. In samples prepared with cleaved-overgrowth techniques, the edge is sharp and 
corresponds to a hard-wall for $x<0$. It is somewhat difficult to model this situation within the present 
formalism, since imposing the hard-wall constraint on the composite fermions does {\it not} impose the 
hard-wall constraint on the electrons; however, we believe that the results presented here have generic validity.

The two principal findings of this work 
are as follows: We find that at filling factor $\nu=1/3$, the edge reconstructions are generic and do not 
depend on the long-ranged nature of interactions, or the smoothness of the confining potential. In contrast 
to this result for $\nu=1/3$, we find that for higher filling factors $\nu=2/5$ and $\nu=3/7$, a 
spin-polarized edge is more robust against reconstruction. The results presented in this paper also show that 
the qualitative behavior of 
the edge profile is the same, irrespective of whether the confining potential is created by a neutralizing 
background which resides in the same plane as the electrons or by a background layer located at a  distance 
$d$ away from the 2D electron layer, as is the case with experimental samples. In real samples, by tilting the 
external magnetic field, the ratio of the Zeeman 
energy to the CF Landau-level splitting can be varied and the $\nu=2/5$ system undergoes a phase-transition 
from a spin-polarized state to a spin-singlet state. In such a singlet state, the region of single-occupancy 
$\rho_{CF}(x)=1$ breaks the spin symmetry. This symmetry breaking and the additional 
energy cost associated with it suggest that the spin-singlet $\nu=2/5$ may undergo reconstruction and 
have spin-textures at the edge. The study of these spin-textures in a spin-singlet state is a topic of 
future work. 

The results in this paper shed some light on the nature of edge states and their dependence on the 
background potential. These results are based on a Hartree-Fock mean-field approximation. The stability of 
these mean-field results can be checked by performing the time-dependent Hartree-Fock analysis of the 
collective excitations at the edge. For example, the onset of reconstruction at $\nu=1/3$ edge can be 
associated with the softening of the edge collective modes at a finite wave-vector.~\cite{wan2,hoang} For 
$\nu=1/3$ a comparison with actual sample parameters shows that the samples are most likely 
in the reconstructed-edge regime. This is a possible explanation~\cite{dhlee,kun} for the disagreement 
between the theoretical prediction~\cite{wen} for the tunneling exponent, $\alpha=\nu^{-1}=3$, and the 
experimental results.~\cite{grayson,chang} At higher filling factors, the effective field theory models, which 
are primarily based on Luttinger liquid description of the edge-modes, lead to various predictions for the 
tunneling exponent $\alpha$ into the edge of a fractional quantum Hall fluid~\cite{ana,ulrich,levitov,dhlee}. 
Similarly, numerical calculations based on wavefunctions show that, in contrast to the CLL prediction, the 
tunneling exponent $\alpha$ may depend on the range of electron-electron interactions~\cite{goldman,mandal}. 
The disagreement between these predictions and recent experiments~\cite{grayson,chang} requires us to re-examine 
the theoretical assumptions regarding edge modes. The formalism presented here provides a concrete method to 
{\it calculate these modes at arbitrary filling factors from a microscopic theory}, and therefore it can be used 
to determine the parameters which characterize various field theory models. To determine the tunneling exponent 
using the extended Hamiltonian theory presents a challenge since, to calculate $\alpha$, it is necessary to 
express electron creation and annihilation operators ({\it not just densities}) in terms of the CF operators. 
This is an open problem and its solution will lead to a microscopic method for the calculation of tunneling 
exponent and other electronic single-particle properties at the edge. Our results at higher filling factors, 
provided that they remain valid at realistic setback distances $d\approx 5l^*$, suggest that the spin-polarized 
edge is more robust against reconstruction, and consequently that the number of edge modes remains unchanged. 
Therefore the resolution of the discrepancy between theory and experiments, if not attributable to edge 
reconstructions, may require new ideas and will deepen our understanding of the edge states in the fractional 
quantum Hall systems. 


\section*{Acknowledgments}
\label{sec: ack}
We wish to thank Kun Yang for helpful comments. This work was supported by the NSF grant DMR-0071611 at the 
University of Kentucky and by the LDRD program at Los Alamos National Laboratory. 


\vspace{3cm}


\end{document}